\documentclass[twocolumn,lp,3p]{elsarticle}

\usepackage{amsmath,amssymb}
\usepackage{times}
\usepackage{graphicx}
\usepackage{natbib}
\usepackage{hyperref}
\usepackage{bm}
\usepackage{slashed}
\usepackage{xcolor}

\newcommand{\bw}{\begin{widetext}}
\newcommand{\ew}{\end{widetext}}
\newcommand{\be}{\begin{equation}}
\newcommand{\ee}{\end{equation}}
\newcommand{\bestar}{\begin{equation*}}
\newcommand{\eestar}{\end{equation*}}

\newcommand{\bi}{\begin{itemize}}
\newcommand{\ei}{\end{itemize}}
\newcommand{\bea}{\begin{eqnarray}}
\newcommand{\eea}{\end{eqnarray}}

\newcommand{\hbo}{\hbox to 1 true cm {\hfill } }

\newcommand{\vc}[1]{\mbox{\boldmath$#1$}}

\newcommand{\ud}{\mathrm{d}}

\newcommand{\m}{{\scriptscriptstyle -}} 
\newcommand{\p}{{\scriptscriptstyle +}}
\newcommand{\LCpm}{{\scriptscriptstyle \pm}}
\newcommand{\LCv}[1]{\mathsf{#1}}

\newcommand{\LCperp}{{\scriptscriptstyle \perp}}

\begin{document}

\begin{frontmatter}

\title{Finite size effects in stimulated laser pair production}

\author[th]{Thomas Heinzl}
\ead{theinzl@plymouth.ac.uk}

\author[ai]{Anton Ilderton}
\ead{anton.ilderton@physics.umu.se}

\author[ai]{Mattias Marklund}
\ead{mattias.marklund@physics.umu.se}

\address[th]{School of Computing and Mathematics, University of Plymouth, Plymouth PL4 8AA, UK}
\address[ai]{Department of Physics, Ume\aa\ University, SE-901 87 Ume\aa, Sweden}

\begin{abstract}
We consider stimulated pair production employing strong-field QED in a high-intensity laser background. In an infinite plane wave, we show that light-cone quasi-momentum can only be transferred to the created pair as a multiple of the laser frequency, i.e.\ by a higher harmonic. This translates into discrete resonance conditions providing the support of the pair creation probability which becomes a delta-comb. These findings corroborate the usual interpretation of multi-photon production of pairs with an effective mass. In a pulse, the momentum transfer is continuous, leading to broadening of the resonances and sub-threshold behaviour. The peaks remain visible as long as the number of cycles per pulse exceeds unity.  The resonance patterns in pulses are analogous to those of a diffraction process based on interference of the produced pairs.
\end{abstract}

\end{frontmatter}


\section{Introduction}
Since Sauter's resolution of Klein's paradox it has been known that the vacuum is unstable to pair production in the presence of a homogeneous electric field exceeding the critical value $E_S = m^2/e \simeq 1.3 \times 10^{18}$ V/m, when the energy acquired by an electron traversing a Compton wavelength equals its rest mass \cite{Sauter:1931zz, Schwinger:1951nm}. The view that the critical field was too large to ever be produced in a laboratory has recently been challenged with the advent of ultra-high power lasers capable of reaching $E\sim 10^{-2}E_S$ \cite{Ringwald:2003iv, Heinzl:2008an}. With plasma tools such as high-harmonic focussing \cite{Gordienko:2005zz} the Sauter-Schwinger limit may come within reach during the next decade.

While the strongest electric fields now available remain below the Schwinger limit, pairs have nevertheless been created in the laser experiment SLAC E-144  \cite{Bamber:1999zt}. Here, the SLAC electron beam was collided with a low intensity optical laser to generate high energy (`probe') photons. These probe photons then combined with the laser to produce electron positron pairs. While nonlinear intensity effects were unambiguously detected, the experiment did not offer a concrete demonstration of effects stemming from the electron mass shift \cite{McD-talk}.

The analysis of the SLAC experiment was based on the infinite plane wave results of \cite{Nikishov:1963,Nikishov:1964a,Narozhnyi:1964}, as the picosecond laser pulse contained around 1000 cycles of beam. In the next generation of experiments, to be performed at facilities such as CLF and ELI \cite{Heinzl:2008an}, laser pulses will have femtosecond duration, corresponding to $\mathcal{O}(1)-\mathcal{O}(10)$ beam cycles. Thus, the use of ultra-short pulses now makes it necessary to go beyond the approximation of infinite pulse extent. It is well known that the pair creation spectra depend very sensitively on the details of the laser field, as is apparent from considering even the simplest constant field models -- for constant electric fields see \cite{Baier:1974hn, Schutzhold:2008pz, Dunne:2009gi, Baier:2009it, Monin:2009aj}, for magnetic fields \cite{Erber:1966vv,Klepikov:1952,Toll:1952, Baier:2007dw} and for crossed fields see \cite{Toll:1952, Heinzl:2006pn}. The impact of pulse shape and duration has recently been extensively investigated in the context of {\it vacuum} pair production \cite{Hebenstreit:2009km, Bulanov:2010ei,Dumlu:2010ua} using purely electric fields which remain spatially homogenous but become time dependent \cite{Brezin:1970xf, MockenNew}. These investigations have revealed a rich {\it substructure} in the spectrum of pairs produced in electric fields, caused by finite beam duration\footnote{For related investigations into nonlinear Compton scattering see \cite{Krafft:2003is, Debus:2009, Heinzl:2009nd, Mackenroth:2010jk} and for ionisation see \cite{ion1, ion2}.}.

Here we extend these finite volume studies to photon--laser collisions, i.e.\ ``stimulated'' pair creation \cite{Schutzhold:2008pz}, using strong-field QED in this particular context for the first time. In contrast to the above approaches we do not neglect the effects of the magnetic field, which are particularly important at high frequencies \cite{Ruf:2009zz}, and therefore model the laser by plane waves of {\it finite} temporal (longitudinal) extent. Note also that the plane wave obeys Maxwell's equations in vacuum, unlike a time dependent electric field.

We briefly address transverse size effects.  One expects the following analysis to be valid provided that the incoming high-energy photons only probe the near-axis region well within the waist size of the high-intensity laser.  With a tight focus (small waist size), as is usually required for high intensity,  the produced pairs would be rather sensitive to the finite transverse extension of the laser beam. On the other hand, a weakly focussed laser will appear approximately as a plane wave (of finite longitudinal extent) to the pairs. In this case, finite transverse size effects should be minor, and the plane wave (or time dependent electric field) model should be a reasonable approximation.  One finds in nonlinear Compton scattering (the crossed process of stimulated pair production) that in the strongly focussed case where the incoming electron beam radius is larger than the laser waist, finite size effects generically lead to broad and unstructured spectra. For a sufficiently weak focus, on the other hand, spectral substructure remains identifiable \cite{Heinzl:2009nd}.

The paper is organised as follows.  We begin with the general expressions for pair production in a plane wave. We then briefly review the well known results for an infinite plane wave (IPW). Comparing these results with pair production in a pulsed plane wave (PPW), we will see how the mass shift arises when the momentum transfer is quantised.  We will also show that the behaviour of the momentum transfer is intimately related to the fine structure of the emission rate, and that pair production in a PPW is essentially a (multiple slit) diffraction process. The emission rates in a PPW exhibit below threshold behaviour relative to the IPW case.
\section{Plane waves and pair production}
We model the laser by a plane wave of frequency $\omega$, i.e.\ a function of the invariant phase $k.x$ ($\omega=|\mathbf{k}|$), with field strength $F_{\mu\nu} = \dot{f}_j(k.x)  ( k_\mu a^j_\nu -a^j_\mu k_\nu )$,  $j\in\{1,2\}$, and a dot is differentiation with respect to $k.x$. The polarisation vectors obey $a^i.a^j=-(e^2a^2/m^2)\delta^{ij}$, defining an invariant, dimensionless amplitude $a$. The associated gauge potential is $A_\mu (k.x) = f_j(k.x) \, a^j_\mu$.

The total pair creation probability can, in principle, be calculated via the (imaginary part of) the polarisation tensor in this plane wave background. The combined space and time dependence makes this a very hard problem, and there has been only limited progress so far \cite{Reiss:1962, becker:1975}. It seems simpler to directly calculate the scattering amplitude for pair production, via strong-field QED using Volkov solutions of the Dirac equation \cite{volkov:1935, Berestetskii:1982}.  The tree level amplitude for production of an $e^\m$ $e^\p$ pair of momenta $p_\mu$ and $p'_\mu$, stimulated by a probe photon of polarisation $\varepsilon$ and momentum $k'$ is
\be \label{SFI}
  S_{fi} = -i e \int \ud^4 x\ \bar{\Psi}^{(\m)}_{p} e^{-ik'.x} \slashed{\varepsilon}\, \Psi^{(\p)}_{p'} \; ,
\ee
where $\bar\Psi^{(-)}_{p}$ and $\Psi^{(+)}_{p'}$ denote the outgoing Volkov electron and positron, respectively. The Volkov electron is
\be \label{VOLKOV}
  \Psi^{(\m)}_p = e^{-iS} \bigg( 1 + \frac{e}{k.p} \slashed{k} \slashed{A} \bigg) u_p \equiv e^{-iS} \Gamma_{\!p}\, u_p \; ,
\ee
where $p$ is its asymptotic momentum,  $u_p$ is a free Dirac spinor and $S$ is the Hamilton--Jacobi action,
\be \label{Q.ACTION0}
  S = p.x - \frac{1}{2k.p} \int\limits_{k.x}^{\infty} \! 2e A.p - e^2 A^2 \equiv p.x + I_p\;.
\ee
Noting that the integrand in (\ref{SFI}) depends nontrivially only on $k.x$,  we introduce light-cone coordinates $x^\LCpm = x^0\pm x^3$, $\vc{x}^\LCperp := (x^1,x^2)$, and take the laser to propagate in the $x^3$ direction so that $k.x\equiv \omega x^\m$. The scattering amplitude then reduces to a light-cone Fourier integral over $k.x$
of $\mathcal{M}(k.x)\equiv\bar{u}_p \overline\Gamma_{p}\,\slashed{\varepsilon}\, \Gamma_{\m p'}v_{p'}\ e^{i(I_p-I_{\m p'})}$, with the light-cone three momenta $\bm{\LCv{p}} := (p_\p, \vc{p}_\LCperp)$ conserved, as follows from (\ref{VOLKOV}) being an eigenfunction of $\bm{\LCv{p}}$. 
Then (\ref{SFI}) becomes
\be \label{FOURIER.MFRAC}
  S_{fi} \!= \!\frac{1}{k_\m}\delta^3 (\bm{\mathsf{p}}' + \bm{\LCv{p}} -\bm{\LCv{k}}')\! \int\! \ud (k.x) \, e^{i(\LCv{y}+\LCv{y}'-\LCv{x}')k.x}{\mathcal{M}}\;,
\ee
where we have introduced three boost invariant light-cone momentum fractions, $\LCv{y}$, $\LCv{y}'$ and $\LCv{x}'$ for the electron, positron and probe photon respectively. They appear through the sum
\be\label{snew}
	 \LCv{y}+\LCv{y}' -\LCv{x}'   \equiv  \frac{p_\m}{k_\m} +  \frac{p'_\m}{k_\m} -  \frac{k'_\m}{k_\m}   \;,
\ee
which we will see contains all the physics, and which we note is Fourier conjugate to the invariant phase $k.x$. We can now look at specific examples.
\section{Pair production in an IPW.}
It is well known that pair production in an IPW is a sum over processes supported on \cite{Nikishov:1963,Nikishov:1964a,Narozhnyi:1964}
\be\label{CONS.INF}
	q_\mu + q'_\mu = n k_\mu + k'_\mu\;.
\ee
Here $n$ is the number of laser photons taken from the beam and $q_\mu$ is the average, or `quasi', momentum of the electron over a laser period which results classically from its rapid quiver motion in the laser field \cite{McDonald:1986zz}. The quasi-momentum may be calculated from the classical orbit, and in an IPW is
\be \label{QUASIMOM.INF}
  q_\mu = p_\mu + \frac{m^2 a_0^2}{2 k.p} \, k_\mu \; ,
\ee
with $a_0$ the dimensionless intensity parameter, often written $ a_0 = eE/\omega m$ in the lab frame, and which equals $a$ ($a/2$) for circularly (linearly) polarised waves \cite{Heinzl:2008rh}. Squaring the quasi-momentum gives the celebrated electron mass shift, $q^2 = m^2 (1 + a_0^2) =:m_*^2$ \cite{Sengupta:1952}. Hence, (\ref{CONS.INF}), summed over $n$, describes the multi--photon production of a {\it heavy} pair with rest masses $m_*$.  It follows from squaring (\ref{CONS.INF}) that a minimum number of laser photons is required, namely the first integer bigger than $s_0^*\equiv 2 m_*^2 / k.k'$. Hence, the effective mass {\it blue}-shifts the energy threshold\footnote{In nonlinear Compton scattering, the mass shift leads to a {\it red}-shift of the Compton edge \cite{Harvey:2009ry}.} from $ 2m^2/k.k'$.

To explain how these structures (including the meaning of a number of photons coming from a {\it classical} background field) arise, we must discuss the relevance of the quasi-momentum for the scattering amplitude (\ref{FOURIER.MFRAC}). Recall that $S=p.x + I_p$ and $I_p$ depends on $k.x\equiv \omega x^\m$. For an oscillatory field the integrand of $I_p$ decomposes into a constant average (or ``light-cone zero mode'' \cite{Heinzl:2003jy}) and a periodic, oscillating fluctuation $\delta I_p$. The average over a cycle is the longitudinal component of the quasi-momentum, so we write
\be \label{Q.ACTION}
 		S = p.x + (q_\m - p_\m)x^\m + \delta I_p\;.
\ee
If we use a bar to denote quasi-momentum fractions, i.e. $\bar{\LCv{y}}\equiv q_\m/k_\m$ instead of ${\LCv{y}} = p_\m/k_\m$, then (\ref{FOURIER.MFRAC}) becomes
\be \label{FOURIER.MFRAC2}
  S_{fi} \!= \!\frac{1}{k_\m}\delta^3 (\bm{\mathsf{p}}' + \bm{\LCv{p}} -\bm{\LCv{k}}')\! \int\! \ud (k.x) \, e^{i(\bar{\LCv{y}}+\bar{\LCv{y}}'-{\LCv{x}}')k.x}{M}\;,
\ee
where $M=\mathcal{M}$ but with $I\to \delta I$.  Now, we have the Fourier integral of a purely oscillatory, periodic function $M$, which may of course be represented as a Fourier \textit{series}, $M(k.x) = \sum_n \tilde{M}_n \, e^{-i n k.x}$. Plugging this into (\ref{FOURIER.MFRAC2}) we finally obtain
\be\label{COMB1}
  S_{fi} =  \frac{1}{k_\m} \delta^3 (\bm{\mathsf{p}}' + \bm{\LCv{p}} -\bm{\LCv{k}}') \sum_n \tilde{M}_n \, \delta (\bar{\LCv{y}}+\bar{\LCv{y}}'-{\LCv{x}}'-n) \; .
\ee
This rather neatly expresses the important and general result that the scattering amplitude is a 4$d$ delta comb, \textit{irrespective} of the detailed functional dependence of the IPW on $k.x$ (as long as it is periodic). The argument of the comb, $\bar{\LCv{y}}+\bar{\LCv{y}}'-{\LCv{x}}'$, contains the longitudinal momentum transfer from the probe photon to the pair, (\ref{snew}), and also the {\it average} transfer from the laser to the pair (\ref{Q.ACTION}) which shifts $p_-$ to $q_-$ ($p'_-$ to $q'_-$). Hence, the delta comb  quantises the total longitudinal momentum transfer to the pair, forcing it to take integer values via $\bar{\LCv{y}}+\bar{\LCv{y}}'-{\LCv{x}}' = n\in\mathbb{Z}$, which is equivalent to
\be\label{LC.RELN}
	q_\m + q'_\m - k'_\m = nk_\m\;.
\ee
Since the IPW quasi-momenta (\ref{QUASIMOM.INF}) differ from the asymptotic momenta only in the minus component (and since $\bm{\LCv{p}}$ is conserved), it follows that (\ref{COMB1}) has support on precisely the sum of delta functions enforcing (\ref{CONS.INF}).  So, the quantisation of longitudinal momentum transfer has the appearance of an integer number of photons being absorbed from the laser and used to create heavy particles of rest mass $m_*$. In an IPW, therefore, the mass shift and photon number are rather closely related. Note that our interpretation, and equation (\ref{LC.RELN}), are consistent with the quantum optics law that photon number and phase are conjugate variables.

The emission rate, calculated from $|S_{fi}|^2$, inherits the delta comb structure of (\ref{COMB1}) with $\tilde{M}_n\to |\tilde{M}_n|^2$, and so becomes an {\it incoherent} sum. The support of the delta function in (\ref{COMB1}) or, equivalently, light-cone momentum conservation (\ref{LC.RELN}), may hence be viewed as (idealised) \textit{resonance} conditions leading to peaks of zero width, for instance, in the triple differential rate, i.e. the rate as a function of $\vc{p}'_\LCperp$ and $n$. Such resonances have been predicted before in discussions of purely time-dependent electric fields \cite{Popov:1973,Narozhny:1973,Mostepanenko:1974im,Popov:2002,Avetissian:2002} which may be viewed as simple models for counter-propagating laser fields. A more realistic version of the latter (taking into account the magnetic field) has recently been studied by numerically solving the Dirac equation. A resonance condition very much akin to (\ref{LC.RELN}) has been found \cite{Ruf:2009zz}.  Clearly, one expects modifications such as line broadening in a more realistic situation, in particular upon taking into account finite pulse duration which is the subject of the following sections.

\section{Pair production in a pulse.}
A pulsed plane wave is described by a field strength $F_{\mu\nu} = F_{\mu\nu} (k.x)$ that goes to zero for sufficiently large modulus $|k.x|$. Typically, it will contain a finite number of cycles, $N$, modulated by a smoothly vanishing envelope function. For the purposes of this paper we take our fields to vanish outside of $k.x\in P \equiv [0, 2\pi N]$. One expects that in the limit of long pulses, i.e.\ large $N \gg 1$, the physics in a PPW approaches that of the IPW. This was the conclusion of Kibble \cite{Kibble:1965zz} who, in the sixties, ended a controversy on the unphysical nature of IPWs by arguing that they were sufficient to describe the long pulses which were then state-of-the-art: the width $\Delta \omega$ of these pulses in frequency space\footnote{Recall that the Fourier transform of a monochromatic IPW is a delta function, corresponding to zero width.} obeyed $\Delta \omega \ll \omega$, which is equivalent to $N \gg 1$. However, modern high intensity lasers typically have pulses that contain only few cycles, $N = O(1)$ such that an IPW can only yield a rather crude idea (at best) of the ongoing physics in this case. In particular, the appearance and interpretation of quasi-momenta and effective mass as averaging effects seem to become questionable.  The extent to which these ideas can be applied to pair production in pulses will be examined in this and the next section.

The simplest ``pulse'' is obtained by retaining periodicity {\it within} $P$, i.e.\ truncating the IPW fields. (Edge effects may be present if the fields do not vanish smoothly, though these should become negligible in the large-volume limit, i.e.\ with increasing $N$.) Thus, we assume for now that $F_{\mu\nu}$ is regularly oscillating for $0 \le x \le  2\pi N$, corresponding to a wave train of $N$ identical cycles \cite{Neville:1971uc}.  We will drop this slightly over-simplistic assumption and add smooth envelopes in due course.

For a PPW the representation (\ref{FOURIER.MFRAC}) of the S--matrix element remains valid, as the only prerequisite there was dependence of the background on $k.x$. Quasi momentum fractions then appear through the decomposition (\ref{Q.ACTION}). For the cutoff pulses in question the quasi-momenta do not necessarily match those of the IPW, but the mass shift typically does. 
Accordingly, we end up with (\ref{FOURIER.MFRAC2}), but with $M$ no longer being strictly periodic. Therefore, performing the phase integral no longer yields a delta comb as in (\ref{COMB1}). Reducing the integral to a sum over single cycles, one may show that (\ref{FOURIER.MFRAC2}) takes the form
\be \label{SINSIN}
  S_{fi} = \frac{1}{k_\m} \delta^3 (\bm{\mathsf{p}}' + \bm{\LCv{p}} -\bm{\LCv{k}}') \frac{\sin N \pi \LCv{z}}{\sin \pi\LCv{z}}\, \tilde{M}_0 \; ,
\ee
where $\tilde{M}_0$ is the contribution from a single cycle, giving an envelope, and we abbreviate the longitudinal momentum transfer as 
\be \label{Z}
  \LCv{z}\equiv \bar{\LCv{y}}+\bar{\LCv{y}}'-\LCv{x}' \; ,
\ee
from here on. Fourier expanding the single cycle contribution to aid the interpretation we find
\be\label{COH}
	\tilde{M}_0 =  \sum_l   M^l_0 \, \text{sinc}\, \pi (  \LCv{z} - l  ).
\ee
Squaring $S_{fi}$, there are no cancellations between terms as in the IPW, and so the emission rate is a {\it coherent} sum, containing \emph{interference} terms. We recognise the product of the $\text{sinc}$ function in (\ref{COH}) and the rapidly oscillating ratio of sines in (\ref{SINSIN}) as the root intensity distribution of light scattering through a finite aperture containing $N$ slits. Thus, pair production is essentially a  diffraction process: finite beam duration corresponds to a finite aperture, while the $N$ cycles of the pulse correspond to $N$ slits. The emission spectrum will therefore take the form of a diffraction pattern, deriving from the interference not of light but of pairs produced at different points in the beam \cite{Lindner}. (Compare the proposal in \cite{King:2010} to perform the double slit experiment by polarising patches of vacuum to create a grating.) Interference is seen through the ratio of sines (\ref{SINSIN}), which has maxima when the momentum fraction transfer is an integer, $\LCv{z}\in\mathbb{Z}$, and $N-2$ additional surrounding maxima of smaller amplitude, resulting in the substructure discussed in the introduction (see also the figures below).

As expected, the peaks signal the persistence of the IPW resonances with the previous delta comb now acquiring finite widths. The longitudinal momentum (fraction) transfer is now continuous, but peaks are still located at $\LCv{z}= n \in \mathbb{Z}$. The resonance condition thus remains valid and one may loosely think of $\LCv{z}$ as something like a continuous photon number. As $\LCv{z}$ involves quasi-momentum fractions, see (\ref{Z}), the peaks still look like the production of mass $m_*$, rather than $m$, pairs. However, the continuous nature of $\LCv{z}$ together with the line broadening phenomenon implies that the sharp IPW cutoff $2m^2_*/k.k'$ becomes washed out as well:  in other words, there is {\it sub-threshold} behaviour! This is completely consistent with Kibble's reasoning since a PPW contains higher frequency components than a monochromatic IPW which obviously lower the threshold. In fact, the reduction is quite significant: the threshold is reduced to $2m^2/k.k'$, independent of the geometry of the pulse.

The threshold reduction can be observed in the detection of pairs at lower energies than the minimum implied by (\ref{CONS.INF}), i.e.\ with energy lower than is required to produce the effective mass or, equivalently, reach the first resonant momentum transfer. To see this, and the preceding, effects explicitly we turn now to the discussion of examples

\section{Examples}

In what follows, we will study PPWs representing a slight generalisation of a model recently suggested by Mackenroth et al. \cite{Mackenroth:2010jk}. We consider a linear polarised wave by setting $f_2 = 0$, and consider the family of gauge potentials
\be \label{PULSE}
  f_1 (k.x) = \left\{ \begin{array}{ll} \sin^K{ \bigg( \displaystyle\frac{k.x}{2N} \bigg)}  \sin (k.x) & k.x \in P \\[3pt]
                                            0 \; , & \mbox{otherwise}
                           \end{array} \right. \; ,
\ee
where the envelope is characterised by an integer $K$. 

\subsection{Example 1: Regular wave train ($K=0$)}
We begin with $K=0$ corresponding to a finite wave train of $N$ cycles \cite{Neville:1971uc} i.e.\ a ``mutilated''  IPW sharply cut off outside of $P$. The potential (\ref{PULSE}) is then continuous, while the field strength is $\dot{f_1} (k.x) = \cos k.x$ inside $P$, zero outside, and therefore has hard (step function) edges. These are unphysical, but we will see how their effects come to disappear in the large-$N$ limit. For this example, the quasi-momenta are as in (\ref{QUASIMOM.INF}), and the mass shift remains the same as in an IPW: $q^2=(1+a^2/2)m^2 = (1+a_0^2)m^2$.
\begin{figure}[t!]
\includegraphics[width=0.95\columnwidth]{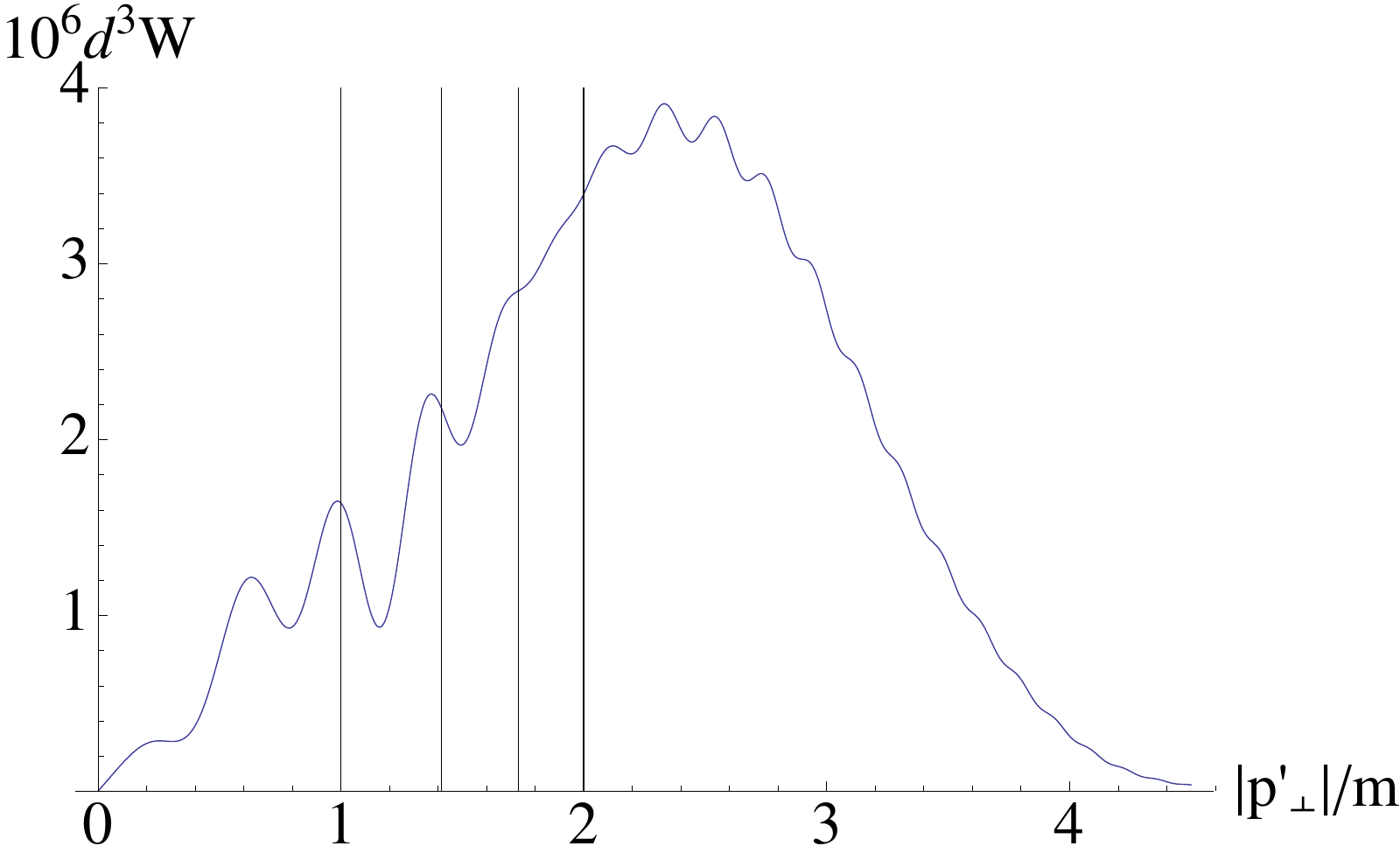}
\includegraphics[width=0.95\columnwidth]{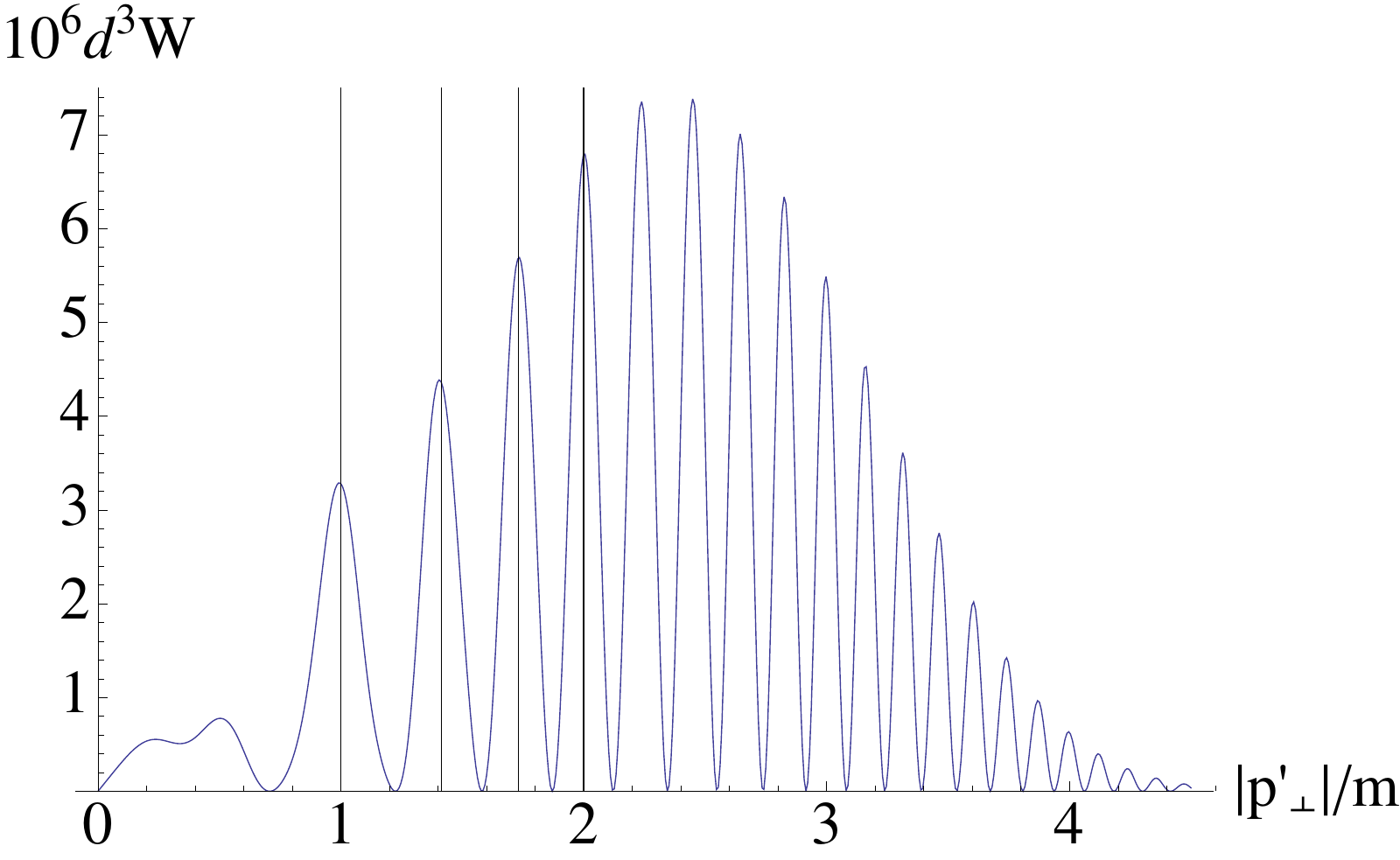}
\includegraphics[width=0.95\columnwidth]{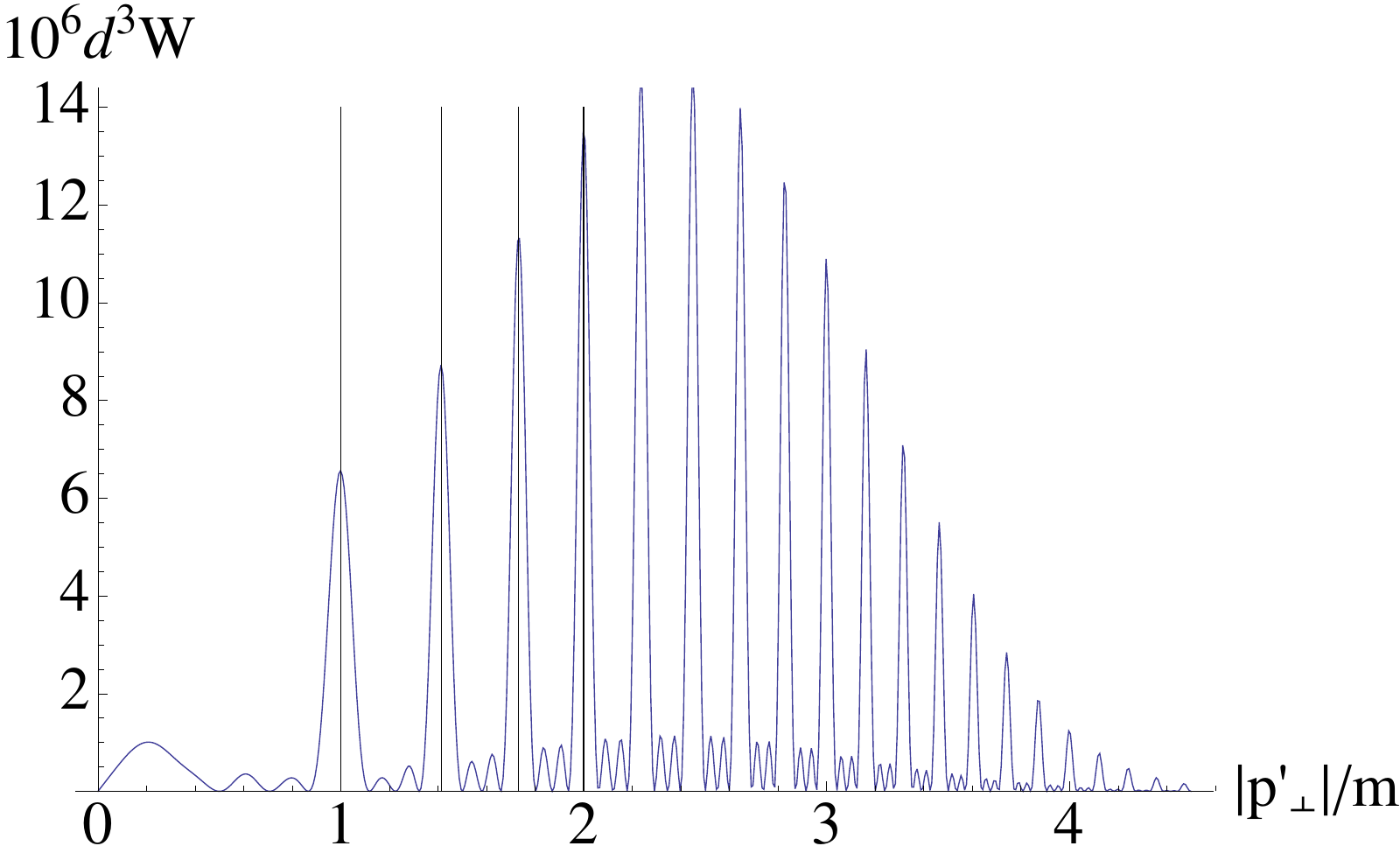}
\caption{\label{3DIFF.FIG}
Triple differential rate, units of $e^2m^2n_\gamma/{32\pi^2\omega'}$, as a function of transverse $e^\p$ momentum. Linear polarisation, $a_0=2$, $\omega'=250$ GeV. $N=1$, $2$ and $4$ cycles of the laser (descending), corresponding to $4\,$fs, $8\,$fs and $16\,$fs pulses respectively.  Black (vertical) lines are the IPW delta comb (quasi-momentum conservation resonances). The non--zero signal to the left of $|\vc{p}_\LCperp'|/m=1$ is production below threshold.}
\end{figure}

Fig.~\ref{3DIFF.FIG}  gives the triple differential rate for pair production in $N=1,2$ and $4$ cycles of the beam, plotted as a function of $|\vc{p}'_\LCperp|/m$ for fixed transverse angle $\phi=0$ and half maximum light-cone component $p'_\p=\omega'/2$.  In accord with the discussion of focussing in the introduction, the intensity is not too high (this is also useful for quick numerical calculation of the rates). Vertical (black) lines are the four lowest terms of the IPW delta comb (the remaining contributions stand to the right).  There is a rich structure even in the single cycle contribution which must be dominated by finite pulse duration, and in this case edge, effects.  As we go beyond one cycle (for which the ratio of sines in (\ref{SINSIN}) is unity) the shape of the differential rate changes significantly. When $N>1$, we have interference effects which are analogous to those appearing when one goes from single to multiple slit diffraction: resonance peaks,  centred on the conservation of quasi-momentum as in (\ref{CONS.INF}), with $N-2$ surrounding subpeaks. As is clear from the properties of the $\sin/\sin$ factor in the rates (and as shown in the plots), the strong peaks dominate in the limit of a large number of cycles, exactly reproducing the IPW delta comb.

We also observe the threshold reduction, through the production of pairs with energy lower than that allowed in the IPW: this is seen as the non--zero signal to the left of the first possible IPW resonance at $|\vc{p}'_\LCperp|/m=1$. Thus, this simple example contains all the promised features. Let us now move on to more physical examples without edge effects.

\subsection{Example 2: Pulses with smooth envelope ($K>1$)} \label{lastsect}
\begin{figure}[t!]
\includegraphics[width=0.9\columnwidth]{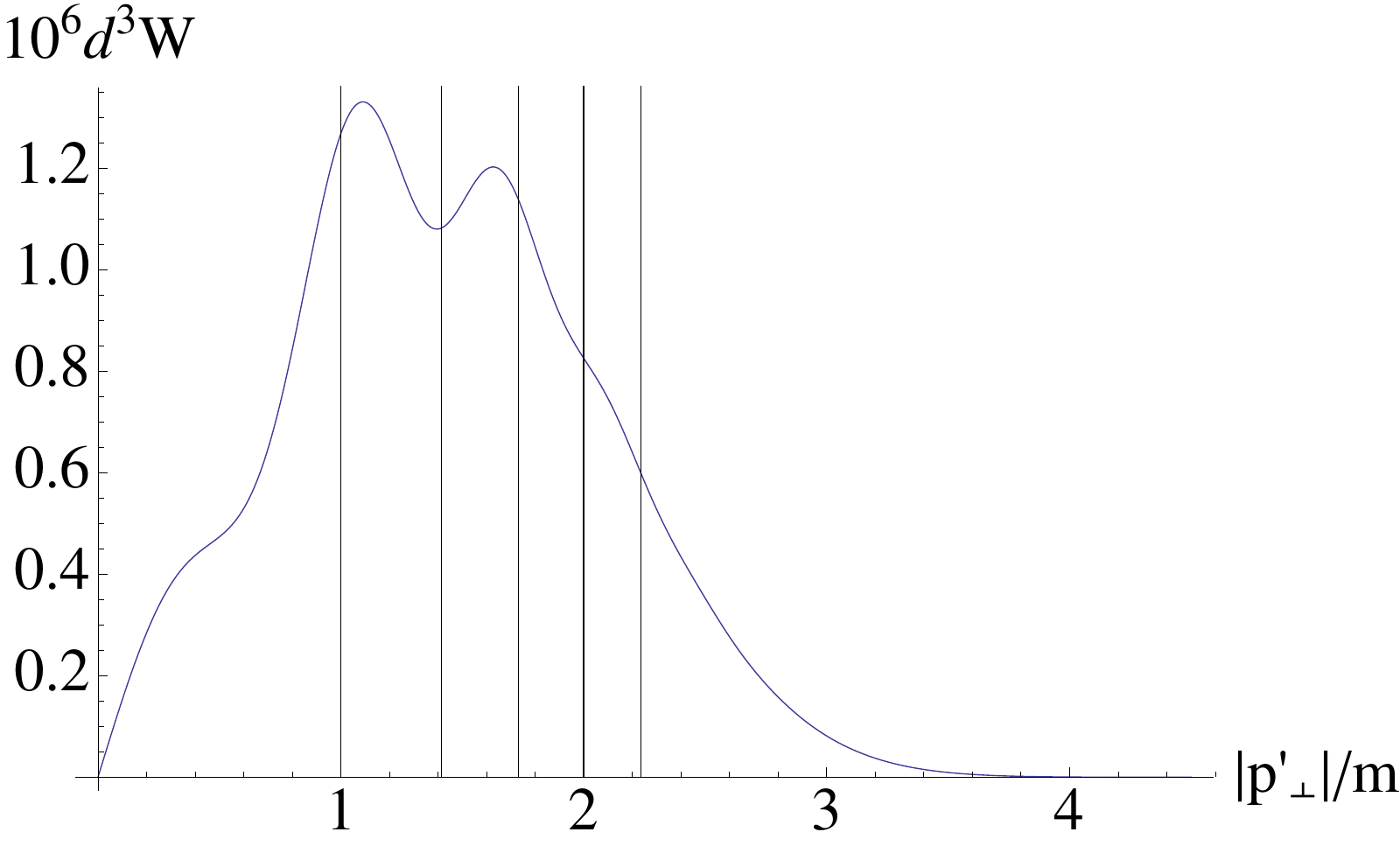}
\includegraphics[width=0.9\columnwidth]{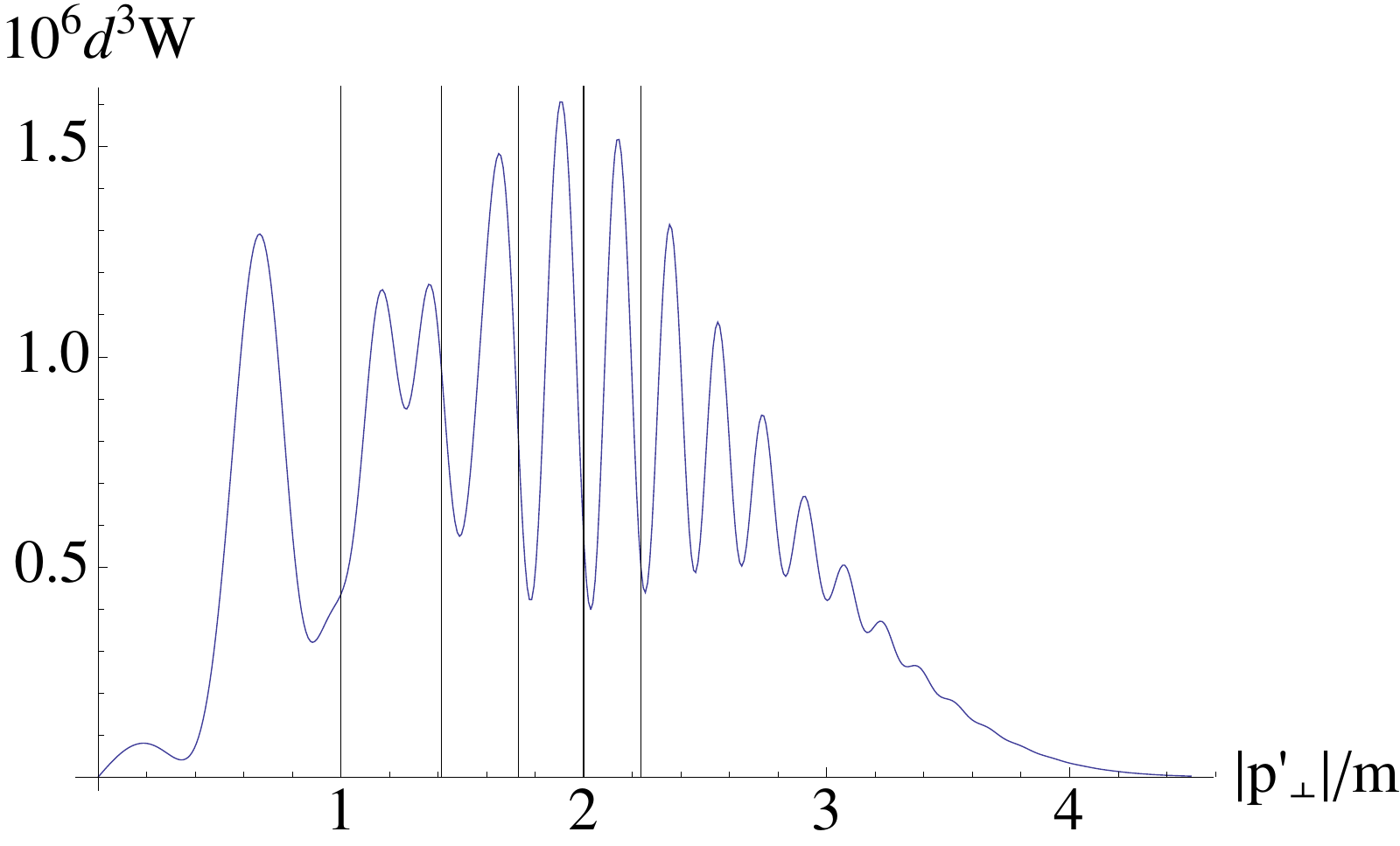}
\includegraphics[width=0.9\columnwidth]{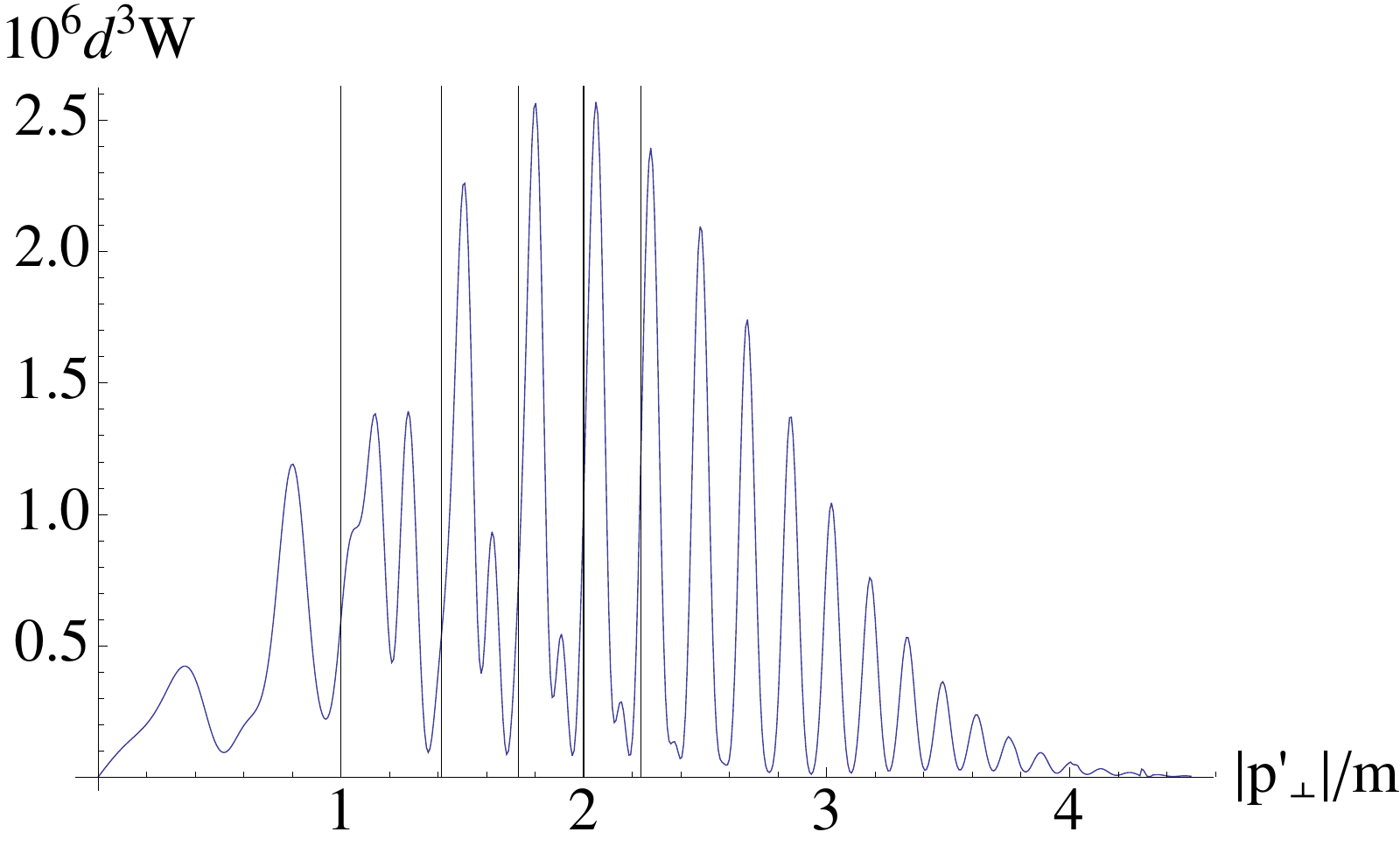}
\caption{\label{3DIFF.FIG-3}
Triple differential cross section for the pulse (\ref{PULSE}), $K=4$, field amplitude $a=2\sqrt{2}$ to match the intensity parameter above. $N=1$, $4$ and $8$ cycles of the laser (descending), corresponding to $4\,$fs, $16\,$fs and $32\,$fs pulses respectively. The diffraction pattern remains. Sub--threshold behaviour remains.}
\end{figure}

\begin{figure}[t!]
\includegraphics[width=0.95\columnwidth]{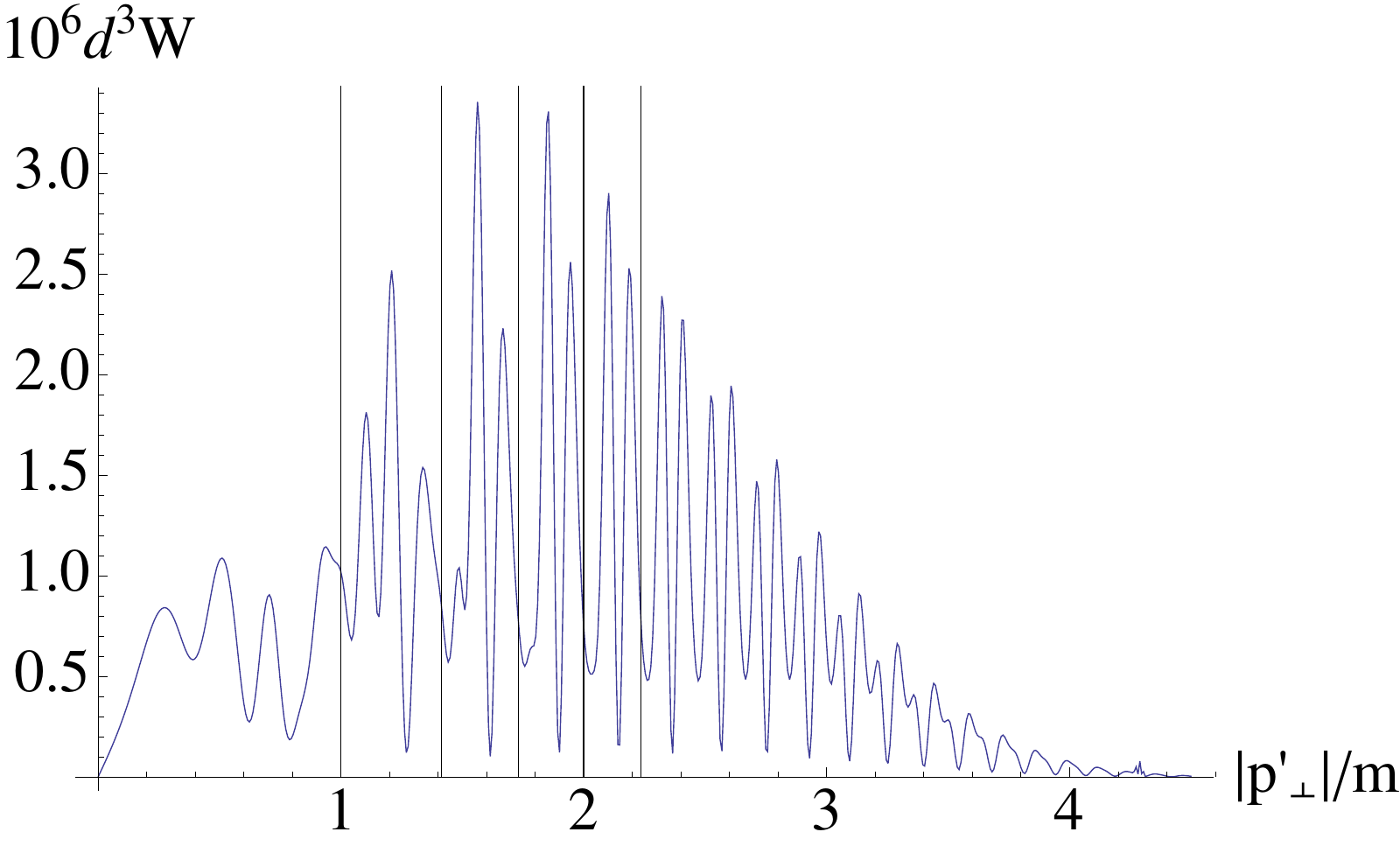}
\caption{\label{QUAD.FIG}
Triple differential cross section for the pulse (\ref{PULSE}) with $K=2$. Other parameters as in Fig.\ \ref{3DIFF.FIG-3} and $32$ fs duration. The rate again exhibits both the diffraction pattern and sub--threshold behaviour.}
\end{figure}

For our second example we add a smooth envelope function. It should be expected from results on vacuum pair production \cite{Hebenstreit:2009km, Dumlu:2010vv} that the resulting spectra will be highly sensitive to the details of this envelope, i.e.\ for our example family (\ref{PULSE}), on the value of $K$. We will see, however, that all the features associated with the diffraction pattern and predicted above continue to appear in the rates. Following \cite{Mackenroth:2010jk} we adopt $K=4$ hence adding a $\sin^4$ envelope to the previous wave train ($K=0$). The field strength $F_{\mu\nu}$ is now {\it smoothly} vanishing at the edges of the pulse. The rate for this pulse is shown in Fig.\ \ref{3DIFF.FIG-3}. Even though the pulse no longer consists of identical cycles, all the features of the diffraction rate remain. (i) There is sub-threshold behaviour relative to the IPW results. (ii) The form of the rate changes dramatically above one cycle of the beam, with rapidly oscillating substructure setting in. (iii) This leads to a series of strong peaks. The envelope in (\ref{PULSE}) naturally smooths out many features of the spectrum, and the fine structure of the diffraction pattern sets in a little slower than in our previous example, so we have plotted $1$, $4$ and $8$ cycles of the laser. Reassuringly, our interpretation of pair production in pulsed plane waves as a diffraction-like process, and the resulting features of the positron emission spectra, are shown to hold for realistic pulses as well. For additional confirmation of this, we display in Fig.\ \ref{QUAD.FIG} the eight-cycle rate for the pulse with the $K=2$ envelope. The field strength is again smooth. The sub-threshold behaviour and interference pattern remain, but  clearly differ from those in Fig.\ \ref{3DIFF.FIG-3}. This illustrates both the expected sensitivity to the pulse envelope, and the robustness of our results.

We remark that in arbitrary pulses, there is no single natural definition of an average from which to construct the quasi-momentum and mass-shift. Assuming \emph{identical} cycles as we did before, the quasi-momenta are given by a cycle average, but do not in general agree with their IPW counterparts (though the mass-shift typically does). Indeed, we notice in the above plots that the strong peaks of the rates do not exactly track the IPW resonances as in the $K=0$ example, which is an indication of (physical) finite size effects affecting the quasi-momentum.  However, our diffraction patterns offers us a potential way to reconstruct the mass shift. This is possible because any separation into an `average' and `fluctuation' of the field will lead to a separation of the Volkov exponent along the lines of (\ref{Q.ACTION}).  This results in an expression for $\LCv{z}$ as a function of $|\mathbf{p}_\LCperp|/m$ (at each angle transverse angle $\phi$) which is just the resonance condition for the given pulse. Hence, by locating the positions of the resonance peaks in the production rates, one could in principle reconstruct this function, and from it the quasi-momenta and the mass shift. We hope to comment further on this idea, and the mass shift in general, in a future paper.


\section{Conclusions.}
We have given a new interpretation and understanding of the electron mass shift in strong-field pair production. We have seen that pair production in a plane wave of finite temporal extent is a diffractive process. The positron emission spectrum can be interpreted as an interference pattern, with a rich substructure. The rate exhibits resonant behaviour when the {\it laser averaged} light-cone momentum transfer to the pair is a multiple of the laser frequency. This resonance condition is equivalent to the momentum conservation of a multi laser-photon (higher harmonic) process which creates a pair of effective mass $m_*$. However, the rates are not completely suppressed away from these conditions, as in the IPW case, and thus the pulsed rates exhibit line broadening and significant below-threshold behaviour, with the electron rest mass, rather than the effective mass, setting the scale. For a large number of cycles the diffraction pattern resembles the delta comb of the IPW limit, centred on the resonant values of momentum transfer. This corresponds to the loss of the coherent (interference) terms in the pair production rates.

Phenomenologically, pair creation certainly requires high energy probes to stimulate the process. With regards to the laser parameters, it seems reasonable to work at moderate $a_0 = O(1)$ which, in turn, allows for longer pulse duration and therefore a large number of cycles per pulse, $N \gg 1$. In this regime, finite pulse duration effects should be under control. This will require some fine tuning for which the present results should provide a solid basis.

Above, we considered the triple differential rates, which, unless the pair yield is very high, would be challenging to measure experimentally. Our focus was on these rates because they are the simplest to plot and understand. It is of course possible to integrate up and thus obtain the double and single differential rates, and the total probability. These will be the subject of a future study.

\subsection*{Acknowledgements}
A.~I. and M.~M. are supported by the European Research Council under Contract No. 204059-QPQV.

\end{document}